\documentstyle[prb,aps,floats]{revtex} 
\begin{document} 
\input epsf
\title{"Electro-flux" effect in superconducting hybrid Aharonov-Bohm rings} 
\author{T. H. Stoof\cite{email} and Yu. V. Nazarov} 
\address{Department of Applied Physics,\\
Delft University of Technology, Lorentzweg 1, 2628 CJ Delft,
The Netherlands}
\address{\mbox{ }}
\address{\parbox{14cm}{\rm \mbox{ }\mbox{ }\mbox{ }
We have extended the circuit theory of Andreev conductance
[Phys.~Rev.~Lett. {\bf 73}, 1420 (1994)] to diffusive
superconducting hybrid structures that contain an Aharonov-Bohm ring.
The electrostatic potential distribution in the system is predicted to be
flux-dependent with a period of the superconducting flux quantum
$\Phi_{0}=h/2e$. When at least one tunnel barrier is present, the
conductance of the system oscillates with the same period.
}}
\twocolumn
\maketitle

Normal metal or semiconductor structures with
superconducting contacts have enjoyed an increasing amount of attention
in recent years. Particularly devices known as Andreev interferometers
have been in the focus of interest.\cite{theory,hekking,circuit,experiment}
The electrical transport in Andreev
interferometers depends on the phase difference of two connected
superconductors, which is a clear manifestation of the coherent nature of
multiple Andreev reflection.\cite{andreev} In two recent
publications\cite{ns,sn} two possible mechanisms were discussed to explain
the experiments of Petrashov et al.\cite{experiment} One of them due to
electron-electron interaction
in the normal metal region and the other related to finite temperatures.
Both mechanisms cause the resistance of such systems to be phase-dependent
with a period of $2\pi$, in contrast to weak localization corrections
to the resistance, which are predicted to display a $\pi$
periodicity.\cite{weakloc} 
 
In the present work we address a different mechanism that causes
an oscillatory resistance in hybrid circuits. This effect does not
depend on the phase difference between to superconducting terminals but
is due to the presence of a magnetic field. If a ring in the normal
metal part of the structure is present, the voltage distribution and
resistance is affected by a magnetic flux through the ring. 
Recently, several experiments along these lines have been
performed.\cite{petrashov}
To study this phenomenon in more detail, we will use a recently developed,
easy-to-use circuit theory of Andreev conductance.\cite{circuit}
With this theory is possible to calculate the 
zero-temperature conductance of diffusive hybrid systems, provided their
size is small enough and the voltages applied are small compared to the
magnitude of the superconducting gap. In this paper we extend
the circuit theory in the of Ref.~\onlinecite{circuit} to account for the
presence of Aharonov-Bohm loops.
We proceed by discussing a novel "electro-flux" effect which is in
principle present in every network that includes an Aharonov-Bohm ring, but
is most pronounced in a circuit consisting solely of diffusive resistors.
Although the conductance is in this case independent of the applied flux,
the electrostatic potential distribution changes periodically with period
$\Phi_{0}=\frac{h}{2e}$. The oscillatory flux-dependence of the
conductance is computed for a few experimentally relevant geometries which
include tunnel junctions.

We consider a diffusive normal metal structure (with diffusion constant
${\cal D}$) connected to one or more superconducting
terminals. The circuit theory of Ref.~\onlinecite{circuit} holds for
sufficiently small systems: $L \ll \xi$
or, equivalently, sufficiently small temperatures and voltages:
$T, V \ll \Delta, {\cal D}/L^{2}$. Here $\xi=\sqrt{{\cal D}/T}$
is the coherence length in the normal metal and $\Delta$ is the magnitude
of the superconducting gap.
Finally we assume that all superconducting terminals are biased at
the same voltage, which allows us to disregard non-stationary
Josephson-like effects.

The theory of Ref.~\onlinecite{circuit} was derived using the
non-equilibrium Green function
technique, originally due to Keldysh\cite{keldysh} and further developed
for superconductivity by Larkin and Ovchinnikov.\cite{larkin}
The basic elements of the theory are the
advanced and retarded Green functions, which determine the energy spectrum
of the quasiparticles, and the Keldysh Green function, which describes
the filling of the spectrum by extra quasiparticles.
At zero temperature, the retarded Green function
$\hat G=s_{x} \hat{\sigma}_{x}+ s_{y} \hat{\sigma}_{y}+s_{z}
\hat{\sigma}_{z}$, where $\hat{\sigma}$ are Pauli matrices, can be
represented by a real spectral vector ${\bf s}=(s_{x},s_{y},s_{z})$.
Due to the normalization of the Green function,\cite{larkin} the
spectral vector is also normalized: ${\bf s}^{2}=1$.
The boundary conditions
on ${\bf s}$ are: ${\bf s}=(0,0,1)$ at all normal terminals and
${\bf s}=(\cos \phi,\sin \phi ,0)$ at all superconducting ones, where
$\phi$ equals the macroscopic phase of the superconducting reservoir.
It is thus possible to map the spatial phase distribution of 
an entire structure on the surface of a hemisphere.

There are two different resistive elements, diffusive resistors and
tunnel junctions. The induced superconductivity in the normal metal
region does not change the diffusive resistance but it does
renormalize the tunnel resistance. The expression for the spectral
current (which is a vector in Pauli-matrix space) through a
resistive element are given by:
\begin{equation}
\label{dif}
R_{\rm D} {\bf I}= \frac{{\bf s}_{1} \times {\bf s}_{2}}{
\sqrt{1-({\bf s}_{1} {\bf s}_{2})^{2}}} \arccos({\bf s}_{1} {\bf s}_{2}),
\end{equation}
for a diffusive resistor with resistance $R_{\rm D}$ and
\begin{equation}
\label{tun}
R_{\rm T} {\bf I}= {\bf s}_{1} \times {\bf s}_{2},
\end{equation}
for a tunnel junction with resistance $R_{\rm T}$. ${\bf s}_{1}$ and
${\bf s}_{2}$ are the spectral vectors on either side of the resistive
element. The circuit-theory rules in terms of the spectral vectors
are:

(i) The Andreev conductance of a system is the same as in normal circuit
theory except for the fact that the tunnel conductivities are
renormalized by a factor ${\bf s}_{1} {\bf s}_{2}$.

(ii) In a normal terminal the spectral vector is the north pole
of the hemisphere whereas in a superconducting one it is located
on the equator, where its longitude $\phi$ indicates the phase
of the superconductor.

(iii) The spectral current is perpendicular to both spectral
vectors on either side of the resistive element. For a diffusive
conductor the magnitude of the current is $I=G_{\rm D} \alpha$ and for
a tunnel junction it is $I=G_{\rm T} \sin\alpha$. Here
$\alpha=\arccos({\bf s}_{1} {\bf s}_{2})$ is the angle between the
two spectral vectors at both ends of the element.

(iv) The {\em vector} spectral current in all nodal points of the network
is conserved.

With these rules it is possible to compute the resistance of a variety
of networks. However, if one wants to include an Aharonov-Bohm ring 
threaded by a flux $\Phi$ into
the circuit, these four rules have to be augmented. To see how this
comes about we perform the standard gauge transformation on the 
Green function to get rid of the explicit vector potential dependence:
\begin{equation}
\label{gauge}
\tilde{G}=\exp({i \chi \hat{\sigma}_{z}}) \;{\hat G}\;
\exp({-i \chi \hat{\sigma}_{z}}),
\end{equation}
where $\chi=\pi \frac{\Phi}{\Phi_{0}}$ and $\Phi_{0}=\frac{h}{2e}$ is
the superconducting flux quantum. In terms of
spectral vectors this gauge transformation is simply a rotation
around the $z$-axis of the original vector by an angle of $2\chi$.
The rotated vector reduces to its original if $2\chi=2\pi$ and
thus will be periodic in the superconducting flux quantum $\Phi_{0}$.
The spectral current vector is rotated likewise.

Without gauge transformation (\ref{gauge}) the equation for $\hat{G}$ would
be rather complicated.\cite{larkin} However, using (\ref{gauge}), the
equation for the transformed Green function $\tilde{G}$ reduces to that for
the original $\hat{G}$ {\it in the absence of flux}.
The flux through the ring now appears in the boundary conditions
on $\tilde{G}$ as follows: At an arbitrary point P in
the ring the Green function $\tilde{G}_{\rm L}$ in P and $\tilde{G}_{\rm R}$
infinitesimally to the right of P are related by:
\begin{equation}
\label{boundary}
\tilde{G}_{\rm L}=\exp({i \chi \hat{\sigma_{z}}}) \;{\tilde G}_{\rm R}\;
\exp({-i \chi \hat{\sigma_{z}}}),
\end{equation}
Hence the spectral vector ${\tilde{\bf s}}_{\rm L}$ is rotated $2\chi$
around the $z$-axis with respect to its 'neighbor' ${\tilde {\bf s}}_{\rm R}$.
Again the spectral current is rotated in the same way. Hence the four
rules remain unaltered but now apply to the transformed $\tilde{\bf s}$
and a fifth rule is needed to prescribe the boundary condition
in the ring:

(v) Going around once in an Aharonov-Bohm ring, the spectral vector at
the end of the loop is rotated by an angle $2\chi$ around the $z$-axis
with respect to the spectral vector at the beginning of the loop.
The same holds for the spectral current.

\begin{figure}[t]
\epsfxsize=7cm \epsfbox[18 280 475 800]{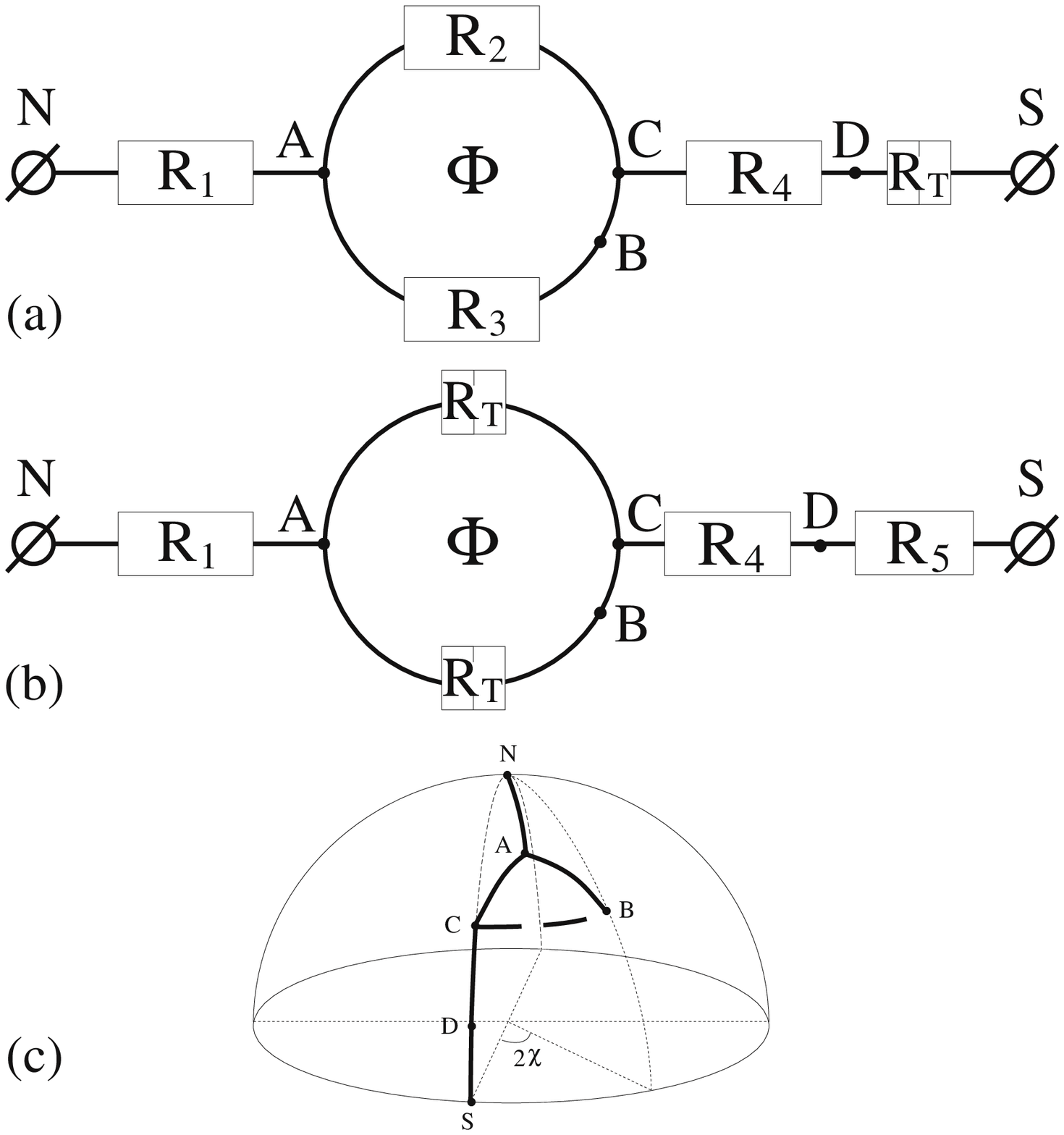}
\caption{(a) and (b) The networks under consideration. $R_{\rm T}$ is a
tunnel junction and all other elements are diffusive resistors.
(c) The circuits mapped onto a hemisphere.} 
\label{fig:system} 
\end{figure}

We now have all the necessary ingredients to calculate the conductance of
the structures depicted in Fig.~\ref{fig:system}. Network (a) consists of
two diffusive wires that connect a normal and a superconducting terminal
to a diffusive Aharonov-Bohm ring. Since a natural place for a tunnel
barrier is at the N-S interface, we also included it in the circuit.
Fig.~\ref{fig:system}(b) shows a SQUID-like device, consisting of a ring
with a tunnel junction in each branch that is connected to the reservoirs
by two diffusive wires.
We consider here a geometry with a single superconducting terminal only
because we want study the effects caused by the applied flux rather than
those due to Andreev interference.
In Fig.~\ref{fig:system}(c) we have mapped the circuits onto a hemisphere
to indicate the position of the spectral vectors. Because only one
superconducting terminal is present, its macroscopic phase is
arbitrary and we choose it to be zero. As can be seen from this
picture we have chosen the point B in the ring as the point where
the spectral vector and current are discontinuous, indicated schematically
by the dashed line.

Since the spectral vectors ${\bf s}_{\rm B}$ and ${\bf s}_{\rm C}$ are
related by Eq.~(\ref{boundary}), we need only compute the positions of the
points A, C and D, which are determined by spectral current conservation
in the nodes (rule (iv)):
\begin{eqnarray}
\label{conserveA}
\sum_{\rm A}{\bf I}&=& {\bf s}_{\rm A} \times \left(
{\bf s}_{\rm N} \frac{C_{\rm AN}}{R_{1}}+{\bf s}_{\rm C}
\frac{C_{\rm AC}}{R_{2}}+{\bf s}_{\rm B}\frac{C_{\rm AB}}{R_{3}}
\right)={\bf 0},\\
\label{conserveC}
\sum_{\rm C}{\bf I}&=& {\bf s}_{\rm C} \times \left(
{\bf s}_{\rm A} \frac{C_{\rm CA}}{R_{2}}+{\bf s}_{\rm D}\frac{C_{\rm CD}}{R_{4}}
\right)+{\bf I}_{\rm BC}={\bf 0},\\
\label{conserveD}
\sum_{\rm D}{\bf I}&=& {\bf s}_{\rm D} \times \left(
{\bf s}_{\rm C} \frac{C_{\rm DC}}{R_{4}}+\frac{{\bf s}_{\rm S}}{R_{\rm T}}
\right)={\bf 0},
\end{eqnarray}
where $C_{\rm IJ}=\arccos({\bf s}_{\rm I} {\bf s}_{\rm J})/
\sqrt{1-({\bf s}_{\rm I} {\bf s}_{\rm J})^{2}}$ and the current ${\bf I}_{\rm BC}=
\Omega \;({\bf s}_{\rm B} \times {\bf s}_{\rm A})\; C_{\rm BA} /R_{3}$, where
$\Omega$ is a matrix that rotates the spectral current over an angle
$-2\chi$ according to rule (v). The physical
current, however, is conserved in every node because a uniform rotation
of the spectral current leaves the physical current invariant.
Note that the spectral current leaving
point A is not equal to the spectral current arriving in C. This is
a consequence of the gauge transformation we have used.
Knowing the spectral vectors in the three points we are able to compute
the resistance of the structure:
\begin{equation}
\label{rest}
R_{\rm tot}=R_{1}+\frac{R_{2} R_{3}}{R_{2} + R_{3}} + R_{4} +
\frac{R_{\rm T}}{\cos \alpha_{\rm DS}},
\end{equation}
where $\cos \alpha_{\rm DS}={\bf s}_{\rm D}{\bf s}_{\rm S}$ is the
renormalization factor for tunnel conductivities according to rule (i).

Let us now first turn to a discussion of what we call the electro-flux
effect. Consider the geometry of
Fig.~\ref{fig:system}(a) without the tunnel junction. In this case the
total resistance of the network is not affected by the applied flux since
the resistance of diffusive elements is not renormalized. However, the
electrostatic potential distribution in the structure is still
flux-dependent. To see this we look at the
zero-temperature expression for the electrostatic potential:\cite{zhou}
\begin{equation}
\label{elpot}
\varphi(x,\Phi)=\frac{1}{4e}\mbox{Tr}\; \hat{G}^{\rm K} = \zeta(x)
\cos \theta(x,\Phi),
\end{equation}
where $\hat{G}^{\rm K}$ is the Keldysh component of the Green function and
$\zeta(x)$ is the quasiparticle distribution function that measures the 
deviation from equilibrium.\cite{circuit} At zero temperature, $\zeta(x)$
is a linear function of position and its slope is proportional to
the voltage drop across a resistive element.
The factor $\cos \theta(x,\Phi)$
in (\ref{elpot}) is just the $z$-component of the spectral vector, which
at zero temperature is equal to the quasiparticle density of states.
From the fact that Eq.~(\ref{elpot}) involves the flux-dependent 
$\theta(x,\Phi)$ it is obvious that also the electrostatic potential
will depend on the flux through the ring. 

\begin{figure}[t]
\epsfysize=7cm \epsfbox[-100 100 250 760]{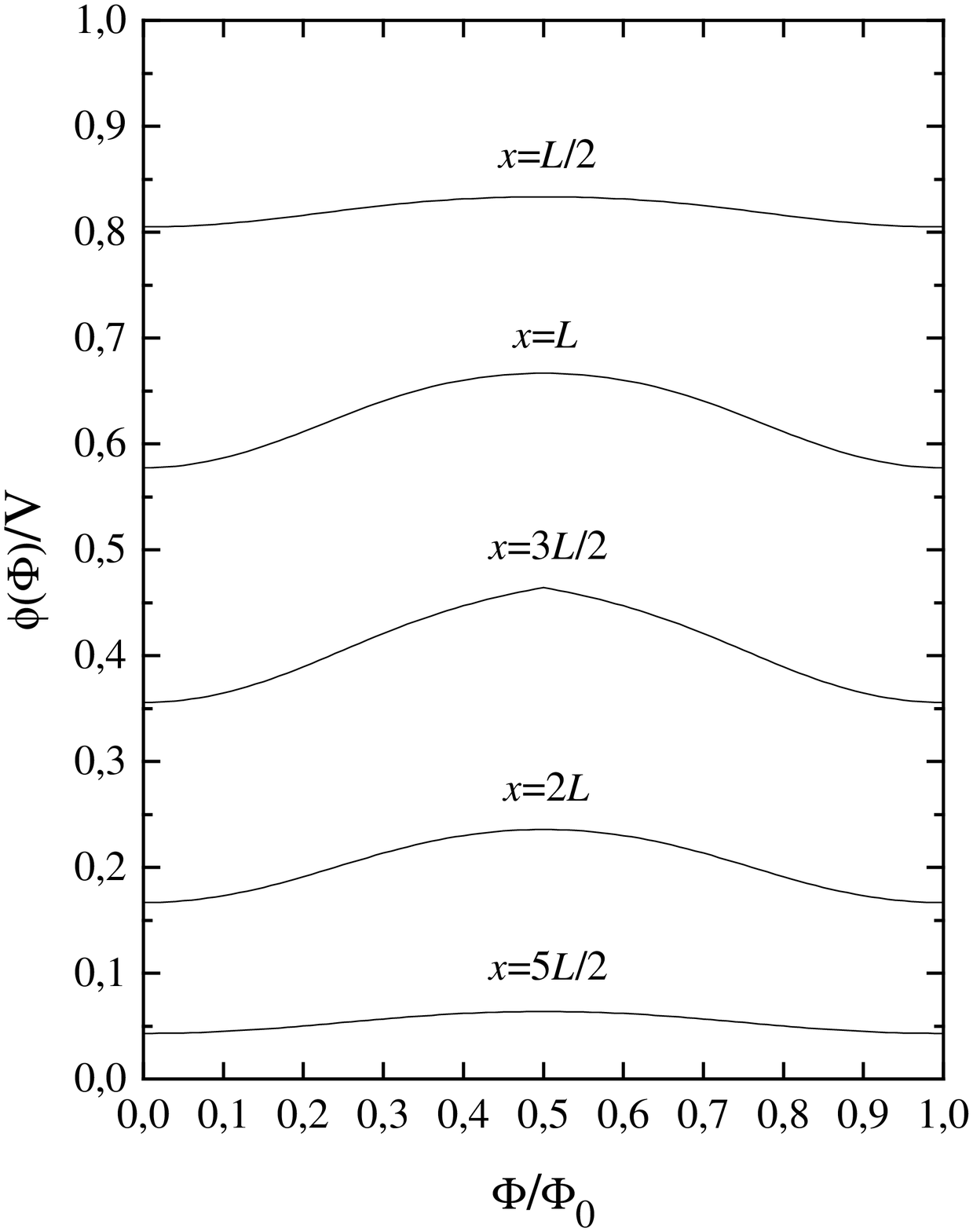}
\caption{Electrostatic potential as a function of flux for different points
along the structure. Calculated for $R_{1} = R_{4} = R$ and
$R_{2} = R_{3}= 2R$.}
\label{fig:potflux} 
\end{figure}

Fig.~\ref{fig:potflux} shows this electro-flux effect at different points
in the structure. 
Here we have considered a structure with a total length of $3L$ and the
wires have lengths NA = AC = CS = $L$.
Fig.~\ref{fig:potflux} clearly shows that the electro-flux effect
is largest in the middle of the structure and vanishes in the end points
of the structure. This new effect is reminiscent of the electrostatic
Aharonov-Bohm effect, in which the phase of an electron in a ring is
influenced by an applied transverse electric field.\cite{elstatab}
However, in a sense the electrostatic Aharonov-Bohm effect is just the
opposite of the electro-flux effect because in the latter case the
electrostatic potential in the ring is modified by changing the phase of
the quasiparticles with a magnetic field.
Using a SET-transistor it should in principle be possible to measure the
local electrostatic potential in a given point. One could then measure the
change in potential as a function of the applied flux. For a more detailed
description of such an experiment see Ref.~\onlinecite{sn}.

\begin{figure}[t]
\epsfxsize=7cm \epsfbox[18 20 450 800]{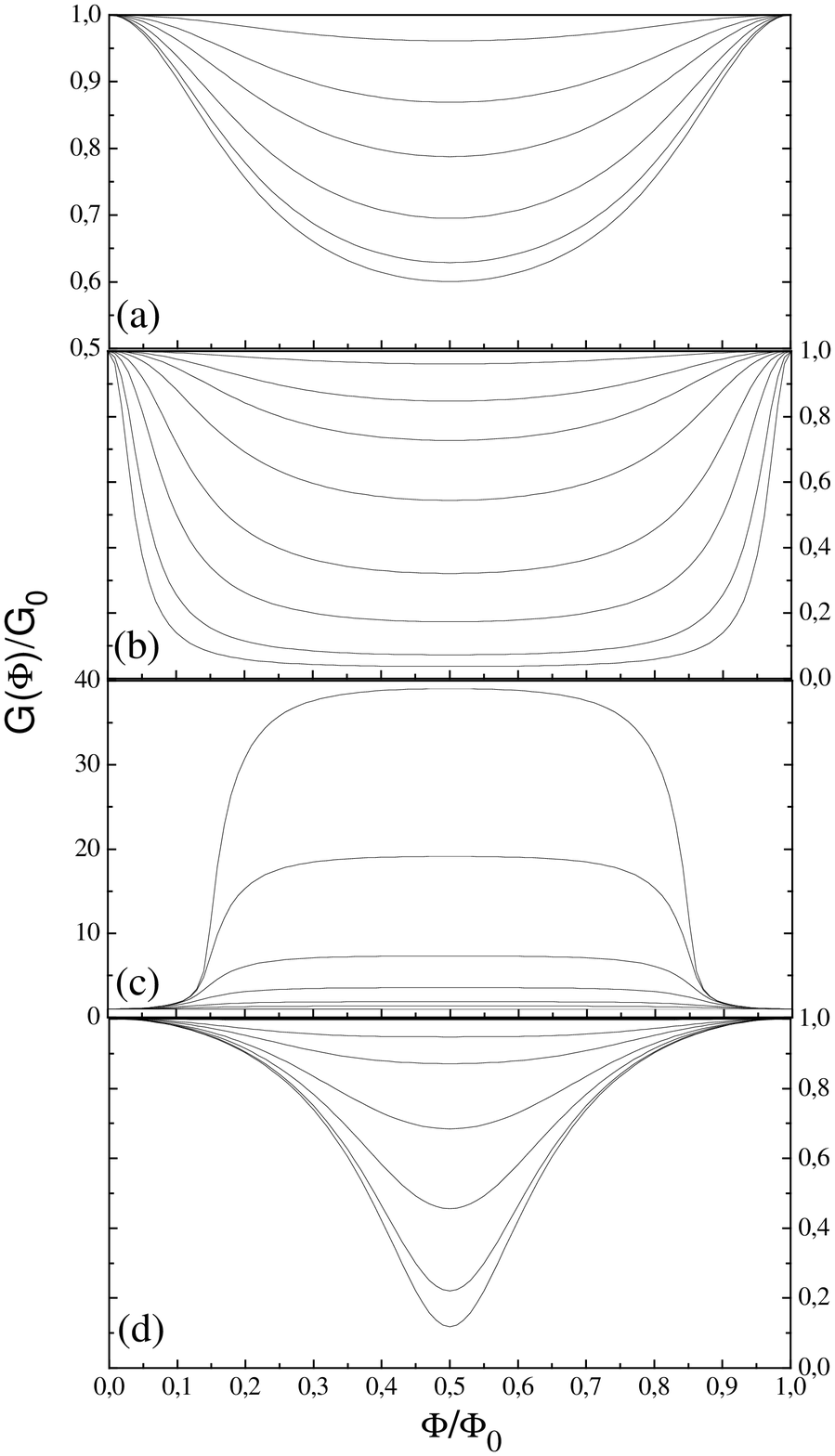}
\caption{Normalized conductance versus applied flux. Panels (a) and (b)
correspond to the structure of Fig.~\protect\ref{fig:system}(a), panel
(c) to the same circuit with $R_{1}$ and $R_{\rm T}$ interchanged and (d)
to the SQUID-like device of Fig.~\protect\ref{fig:system}(b).
From small to large amplitude the different curves correspond to:
(a) $R_{\rm T}/R$=1, 2, 3, 5, 10, 100.
(b) $R_{\rm T}/R$=1, 2, 3, 5, 10, 20, 100.
(c) $R_{\rm T}/R$=1, 3, 5, 10, 20, 50, 100.
(d) $R_{\rm T}/R$=3, 5, 10, 20, 50, 100.} 
\label{fig:condflux} 
\end{figure}

In the last part of this paper we discuss the flux-dependent conductance
of several circuits that may be experimentally relevant. 
In Fig.~\ref{fig:condflux} we have plotted the conductance of three
different systems as a function of the applied flux for different values of
the resistances in the circuit. The conductance has been normalized to its
zero-flux value.
Panels (a) and (b) show the results for the system of
Fig.~\ref{fig:system}(a). In panel (a) $R_{1}=R_{2}=R_{3}=R_{4}=R$ and in
panel (b) $R_{2}=R_{3}=R_{4}=R$ and $R_{1}=R_{\rm T}$. The different curves
correspond to different values of $R_{\rm T}/R$.
Fig.~\ref{fig:condflux}(c) shows the case in which the diffusive 
resistor $R_{1}$ and the tunnel barrier $R_{\rm T}$
in Fig.~\ref{fig:system}(a) have been interchanged and the remaining panel
shows the conductance of the SQUID-like device of Fig.~\ref{fig:system}(b)
with $R_{4}=R_{5}=R$ and $R_{1}=R_{\rm T}$.

Let us first consider the circuit of Fig.~\ref{fig:system}(a).
The panels (a) and (b) show that, in this case, applying a flux through the
ring decreases the conductance. This is easily understood with the aid of
Fig.~\ref{fig:system}(c). When a flux is applied, all points A, C and D
are 'pulled' towards the north pole of the hemisphere, thus increasing the
angle $\alpha_{\rm DS}$ between the spectral vectors ${\bf s}_{\rm D}$
and ${\bf s}_{\rm S}$. Eq.~(\ref{rest}) then shows that
this increases the resistance relative to the zero-flux value. 

It is also clear that an increase of the resistance of both the tunnel
junction $R_{\rm T}$ and the interface resistance $R_{1}$ causes a bigger
effect on the conductance. This because in case (a) point D is much
closer to the north pole of the structure than in case (b), where it is
somewhere in the middle between N and S. Applying a flux will have a
much larger effect on the renormalization factor $\cos \alpha_{\rm DS}$ in
case (b) than in case (a). Whereas the maximal reduction in conductance in
case (a) is less than a factor of 2, it is almost a factor of 30 in case
(b). In the limit of large $R_{1}$ and $R_{\rm T}$ our results agree
with those obtained in Ref.~\onlinecite{hekking}.

Although Fig.~\ref{fig:condflux}(a) and (b) might give the impression that
the conductance always decreases when a flux is present, this is not
generally the case. It is also possible to increase it, e.g.
in systems with a single tunnel barrier between the normal contact and the
ring. When all diffusive resistors are kept constant and only the tunnel
resistance is varied, the conductance is increased dramatically.
As shown in Fig.~\ref{fig:condflux}(c) the maximum increase is a factor of
40 for a tunnel barrier that has a 100 times bigger resistance
than the diffusive resistors in the network.

In the SQUID-like structure of Fig.~\ref{fig:system}(b), the results are
qualitatively the same as those shown in Fig.~\ref{fig:condflux}(b).
Similar considerations as the ones used above 
show that the conductance reduction is largest when both the resistance
$R_{1}$ and the tunnel resistances in the ring are large. There is, however,
a striking difference in shape of the curves. Whereas in panel
\ref{fig:condflux}(b) the 
minimum becomes broader on increasing the resistances, the opposite is
occurring in panel \ref{fig:condflux}(d) where a sharp peak develops.
The characteristic shapes of the curves displayed in Fig.~\ref{fig:condflux}
should be observable experimentally. 

In conclusion, we have generalized the circuit theory of Andreev conductance
of Ref.~\onlinecite{circuit} to networks that include an Aharonov-Bohm
ring penetrated by a magnetic flux. We have given the complete set of
altered circuit-theory rules and used them to calculate the flux-dependent
resistance of several experimentally relevant structures. Under the
right conditions these devices are very sensitive to the applied flux.
We have predicted an electro-flux effect in these circuits, which entails
that the electrostatic potential distribution in the structure can be
altered by varying the applied magnetic flux through the ring. It should be
possible to observe this effect experimentally.

It is a pleasure to acknowledge useful discussions with Michel Devoret,
Daniel Est\`eve, Gerrit Bauer, Mark Visscher and Luuk Mur.


\begin{references}
\bibitem[\protect\dag]{email}Electronic address: theo@duttnto.tn.tudelft.nl
\bibitem{theory} See e.g. H. Nakano and H. Takayanagi, Sol. St. Comm.
{\bf 80}, 997 (1991); A. V. Zaitsev, Phys. Lett. A {\bf 194}, 315 (1994);
A. Kadigrobov et al. Phys. Rev. B {\bf 52}, 8662 (1995).
\bibitem{hekking} F. W. J. Hekking and Yu. V. Nazarov, Phys. Rev. Lett.
{\bf 71}, 1625 (1993)
\bibitem{circuit} Yu. V. Nazarov, Phys. Rev. Lett. {\bf 73}, 1420 (1994).
\bibitem{experiment} P. G. N. de Vegvar et al., Phys. Rev. Lett. {\bf 73},
1416 (1994); H. Pothier et al., Phys. Rev. Lett. {\bf 73}, 2488 (1994);
A. Dimoulas et al. Phys. Rev. Lett. {\bf 74}, 602 (1995);
V. T. Petrashov et al, Phys. Rev. Lett. {\bf 74}, 5268 (1995).
\bibitem{andreev} 
A. F. Andreev, Sov. Phys. JETP {\bf 19}, 1228 (1964); {\bf 24}, 1019
(1967). 
\bibitem{ns} Yu. V. Nazarov and T. H. Stoof, Phys. Rev. Lett. {\bf 76},
823 (1996).
\bibitem{sn} T. H. Stoof and Yu. V. Nazarov, to appear in Phys. Rev. B.
(June 1996).
\bibitem{weakloc} B. Z. Spivak and D. E. Khmelnitskii, JETP Lett. {\bf 35},
412 (1982) [Pis'ma Zh. Eksp. Teor. Fiz. {\bf 35}, 334 (1982)];
B. L. Altshuler, D. E. Khmelnitskii, and B. Z. Spivak, Sol. St. Comm.
{\bf 48}, 841 (1983).
\bibitem{petrashov} V. T. Petrashov et al., Phys. Rev. Lett. {\bf 70},
347 (1993); V. T. Petrashov et al., JETP Lett. {\bf 60}, 606 (1994)
[Pis'ma Zh. Eksp. Teor. Fiz. {\bf 60}, 589 (1994)]. H. Courtois et al.,
Phys. Rev. Lett. {\bf 76}, 130 (1996). S. G. den Hartog et al., preprint.
\bibitem{keldysh} L. V. Keldysh, Sov. Phys. JETP {\bf 20}, 1018 (1964)
[Zh. Eksp. Teor. Fiz. {\bf 47}, 1515 (1964)].
\bibitem{larkin} A. I. Larkin and Yu. V. Ovchinnikov,
Sov. Phys. JETP {\bf 41},
960 (1975) [Zh. Eksp. Teor. Fiz. {\bf 68}, 1915 (1975)]; Sov. Phys. JETP
{\bf 46}, 155 (1977) [Zh. Eksp. Teor. Fiz. {\bf 73}, 299 (1977)].
\bibitem{zhou} F. Zhou, B. Spivak, and A. Zyuzin, Phys. Rev. B {\bf 52},
4467 (1995); See also Ref.~\onlinecite{sn} and references therein.
\bibitem{elstatab} T. H. Boyer, Phys. Rev. D {\bf 8}, 1679 (1973);
S. Datta et al., Appl. Phys. Lett. {\bf 48}, 487 (1986);
S. Washburn et al., Phys. Rev. Lett. {\bf 59}, 1791 (1987).
\end{references}
\end{document}